\documentclass[12pt,final,onecolumn]{IEEEtran}

\usepackage{hyperref}
\usepackage{graphicx}
\usepackage{cancel}
\usepackage{bm}
\usepackage{amsmath}
\usepackage{amsfonts}
\usepackage{booktabs}
\usepackage{caption,booktabs}
\usepackage{lmodern}	
\usepackage[ruled,vlined]{algorithm2e}

\DeclareMathOperator*{\minimize}{minimize}

\begin{document}
%
\title{Iterative Best Response for Multi-Body Asset-Guarding Games}
%
%
%

\author{Emmanuel~Sin, Murat~Arcak, Douglas~Philbrick, Peter~Seiler
\thanks{E. Sin is with the Department of Mechanical Engineering, University of California, Berkeley, CA.}
\thanks{M. Arcak is with the Department of Electrical Engineering and Computer Sciences, University of California, Berkeley, CA.}
\thanks{D. Philbrick is with the Naval Air Warfare Center Weapons Division, U.S. Naval Air Systems Command, China Lake, CA.}
\thanks{P. Seiler  is with the Department of Electrical Engineering and Computer Science, University of Michigan, Ann Arbor, MI.}
\thanks{Respective email addresses: (\href{mailto:emansin@berkeley.edu}{emansin@berkeley.edu}), (\href{mailto:arcak@berkeley.edu}{arcak@berkeley.edu}), (\href{mailto:douglas.philbrick@navy.mil}{douglas.philbrick@navy.mil}), (\href{mailto:pseiler@umich.edu}{pseiler@umich.edu}) }
}

\maketitle

\begin{abstract}
We present a numerical approach to finding optimal trajectories for players in a multi-body, asset-guarding game with nonlinear dynamics and non-convex constraints. Using the Iterative Best Response (IBR) scheme, we solve for each player's optimal strategy assuming the other players' trajectories are known and fixed. Leveraging recent advances in Sequential Convex Programming (SCP), we use SCP as a subroutine within the IBR algorithm to efficiently solve an approximation of each player's constrained trajectory optimization problem. We apply the approach to an asset-guarding game example involving multiple pursuers and a single evader (i.e., $n$-versus-$1$ engagements). Resulting evader trajectories are tested in simulation to verify successful evasion against pursuers using conventional intercept guidance laws.
\end{abstract}


%
\IEEEpeerreviewmaketitle

\section{Introduction}

In his seminal work on differential games, Isaacs describes a pursuit-evasion game that he calls \textit{Guarding a target}  \cite{Isaacs}. In this game, the motive of the pursuer $P$ is to guard an asset $C$ from an attack by the evader $E$. The motive of $E$ is to reach its target while evading $P$. The payoff in this zero-sum game is the distance between $C$ and the point of $P$'s capture of $E$. Assuming simple motion (i.e., constant velocity with the ability to change heading angle instantaneously), the optimal trajectories of $P$ and $E$ are depicted and explained using geometric arguments alone. But when we attempt to build upon this basic game by introducing more complex motion, imposing state and input constraints, or adding more players with non-diametrically opposed objectives, arriving at such analytical solutions becomes difficult. 
\begin{figure}[!htbp]
\centering
\includegraphics[width=0.9\textwidth]{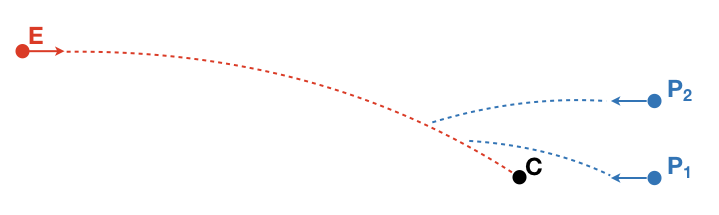}
\caption{Asset-guarding game with Evader $E$, Pursuers $P$, and Asset $C$}
\label{fig:introfigure}
\end{figure}

In this work we present a framework to model advanced forms of the asset-guarding game and introduce a numerical solution method to find optimal strategies for the players. In our version of Isaacs' guarding game we have an asset $C$, a group of $n$ pursuers that we collectively call $P = \{P_1, P_2, \ldots, P_n\}$, and a single evader $E$. The asset $C$ can be specified as a point, surface, or volume to represent, for example, a vehicle, an area of land, or a large structure. We assume that its motion follows a known trajectory (e.g., stationary or moving with constant velocity). The evader $E$ and pursuers $P$ are always active, strategic players in our game. Depending on the application at hand, we may approach the game from the perspective of either $E$ or $P$. If we approach the problem from the perspective of $E$, we are concerned with finding a dynamically-feasible trajectory that both reaches $C$ and avoids capture by $P$ during the course of the trajectory. This reach-while-avoiding condition is non-negotiable in $E$'s mission and so we consider it as a constraint that must be satisfied. On the other hand, the specific final time at which $E$ captures $C$ is negotiable - a later arrival time may be acceptable if $E$ is required to perform evasive maneuvers. Hence, it is natural for $E$ to define its objective as the minimum time required to fulfill its mission-critical constraints. $E$'s speed and acceleration capabilities are greater than those of $C$, but they may be inferior to those of $P$. In any case, $E$ must contend with the sheer number of pursuers that comprise $P$. If found, a trajectory for $E$ that can evade numerous, superior pursuers would be remarkable.

If we now switch our perspective to consider the game from $P$'s point-of-view, we are a group of pursuers, each engineered with the task of intercepting a maneuvering target. If we assume that it only takes one pursuer to disable $E$, we may employ the other pursuers to assist, perhaps by corralling $E$ in a certain direction. This approach can lead to interesting collaborative behavior but it may also result in our use of pursuer vehicles in a manner other than which they were designed for (e.g., using interceptor missiles with lethal warheads simply as herders). Hence, we assume that each pursuer should be programmed with the intent to intercept $E$, without regard to objective formulations for other pursuers. A practical advantage of this assumption is that if a pursuer fails during the engagement, other pursuers will still be on feasible trajectories to intercept $E$. An important temporal constraint on $P$ is that each pursuer's intended intercept time of $E$ should be less than $E$'s predicted intercept time of $C$. The pursuer that can intercept $E$ in the shortest amount of time decides the performance of $P$ as a whole. An interesting situation to consider is the case where $P$'s maneuverability is significantly inferior to that of $E$ - to see if quantity can outperform quality. This discussion of the players' objectives and constraints are mathematically expressed as optimal control problems in a following \textit{Problem Statement} section. 

\subsection{Relevant Work}

Isaacs initiated the study of differential games through a series of unpublished RAND Corporation reports \cite{Isaacs0} starting in 1954. Isaacs' approach resembled that of dynamic programming where he derived control laws for the players and conditions for certain outcomes in simple motion games. Soon after Isaacs' initial reports, Berkowitz and Fleming introduced a calculus of variations approach where they constructed necessary and sufficient conditions for a saddle-point that must be satisfied along the trajectories of the players \cite{BerkovitzFleming}. The approach is applied to a wide class of differential games in \cite{Berkovitz} but did not treat specific examples or applications. In 1965, Ho, Bryson and Baron used the same variational techniques to derive the well-known proportional navigation intercept guidance law by modeling a pursuit-evasion game \cite{Ho}.

We note two state-of-the-art numerical methods for finding local solutions in a wide-class of differential games involving multiple players, non-zero-sum objectives, nonlinear dynamics and constraints. The first method \cite{Fridovich} is inspired by the iterative linear quadratic regulator (iLQR) approach used for optimal control of nonlinear dynamical systems \cite{Li}. However, rather than using the closed-form solution to the LQR problem, the closed-form solution to the n-player, non-zero-sum, linear-quadratic dynamic game \cite{Basar1} is used at each iteration of the algorithm. State and input constraints are implemented as soft constraints, penalizing constraint violations in the objective. The second method \cite{Manchester} also approaches dynamic games by iteratively solving linear-quadratic approximations of the problem. Constraint satisfaction is encouraged by introducing penalty terms in the players' objectives to form augmented Lagrangians. Since a first-order condition of optimality for each player requires that the gradient of its augmented Lagrangian be null at an optimal point, a root-finding problem is formulated and solved using Newton's method. Both methods require initial guesses for all players - either initial feedback control strategies or initial trajectories for states, inputs, and Lagrange multipliers associated with any constraints. 

Yet another solution method is inspired by the concept of \textit{best response}, used in normal-form games \cite{Osborne}-\cite{Nisan}, where the optimal strategy for each player is determined by assuming strategies of all other players are known and fixed. The concept has been used to find open-loop Nash equilibria for simple finite-time horizon differential games in \cite{Dockner}-\cite{Leitmann}. Recent work has used the best response concept to numerically find Nash equilibria in practical examples involving autonomous car trajectory planning and competitive vehicle racing \cite{Fisac1}-\cite{Liniger}. A common theme is to employ an \textit{Iterative Best Response} (IBR) scheme where players' strategies are sequentially and iteratively updated until all players have converged on their best responses. One drawback of IBR, as pointed out in \cite{Fridovich} and \cite{Manchester}, is that the application of existing trajectory optimization methods in an IBR framework may be computationally inefficient with long solution times, detracting from its practical use in differential games.

Recent advances in sequential convex programming (SCP) have enabled efficient, real-time trajectory optimization for constrained, nonlinear systems. SCP is an iterative method that repeatedly formulates and solves a convex, finite-dimensional parameter optimization problem that approximates the original non-convex optimal control problem. A convex formulation is typically achieved by linearizing the nonlinear system around a nominal trajectory (i.e., the solution from the previous iteration) and approximating any non-convex constraints with first-order or second-order approximations. In this convex form, fast and reliable Interior Point Method algorithms \cite{Nocedal} may be used as solvers. Successive Convexification, a type of sequential convex programming, is used in \cite{Szmuk0} to solve a minimum-fuel, 6-DOF powered descent Mars landing guidance problem. In \cite{Szmuk1}, a time-normalization procedure allows for the treatment of minimum-time problems. After these initial works that laid the foundation for SCP, subsequent work by the authors and collaborators aimed at improving various aspects of the method. For example, the issue of guaranteed constraint satisfaction in-between temporal nodes of a transcribed problem is addressed and applied to obstacle avoidance constraints \cite{Dueri}. Given the multitude of options for discretizing continuous-time systems, various methods are compared in terms of accuracy and efficiency in \cite{Malyuta0}. The tutorial \cite{Mao0} places successive convexification in the context of trajectory planning and tracking for aerospace applications and the convergence properties of the method are studied in \cite{Mao1}. Then, the method's applicability is broadened to problems involving mixed integer constraints \cite{Reynolds0}-\cite{Malyuta1}. Most recently, \cite{Szmuk3}-\cite{Szmuk4} focus on real-time implementations with computation times within fractions of a second.

Even without real-time SCP implementation, the IBR scheme serves as a valuable tool when analyzing certain types of differential games. For example, in our asset-guarding game we are willing to assume that the evader $E$ acts first and the pursuers $P$ respond in turn --- without an evader, there is no need for pursuers. If $E$ finds a solution to its optimal control problem (i.e., a feasible trajectory and corresponding strategy) that forces $P$'s optimal control problem to become infeasible, then $E$ may as well announce its strategy in advance and execute it in open-loop. In a deterministic setting, $E$ has no incentive to change its trajectory as the game evolves since $P$ cannot feasibly respond to its current strategy. Knowing whether such a strategy exists is of obvious value to $E$ but is also important to $P$ and $C$, who may decide to recruit more pursuers or strategically change initial conditions. Such an analysis can be carried out at the beginning of an asset-guarding engagement and not necessarily executed online. \\

\subsection{Main Contribution}

The main contributions of this work are:
\begin{enumerate}
\item Formulation of optimal control problems (OCPs) that capture the nature of the asset-guarding game for the Asset $C$, Pursuers $P$, and Evader $E$. The OCPs allow for nonlinear dynamics and non-convex constraints that arise in multi-body, asset-guarding games involving aerospace vehicles.
\item Demonstration of an Iterative Best Response (IBR) solution method that uses SCP trajectory optimization as a subroutine to find solutions for players in a differential game. In our numerical examples, we demonstrate the ability to apply SCP to dynamical models with data in tabular form. Since many practical aerospace systems require the use of lookup tables (e.g., atmospheric parameters, aerodynamic coefficients), this technique may further the adoption of SCP as a powerful trajectory optimization for aerospace systems.
\end{enumerate}
Together, the main contributions provide a framework to model differential games, such as the asset-guarding game, and a solution method to find trajectories that satisfy the players' objectives, constraints, and dynamics. The combination of the IBR algorithm and SCP subroutine provides a practical tool to design and analyze strategies for dynamical systems in conflict. 

\subsection{Notation}

Bold, lower-case letters are used to indicate (column) vector-valued quantities while unbolded letters refer to scalar quantities. The letters $\mathbf{x}$, $\mathbf{u}$, and $t$  are reserved to denote state, input, and time trajectories, respectively. We use $t_1$ for the initial time and $T$ for the final time of a problem. Subscript indices are used in shorthand notation to refer the state and input at specified temporal nodes of a trajectory, e.g., $\mathbf{x}_1 := \mathbf{x}(t_1)$, $\mathbf{x}_k := \mathbf{x}(t_k)$, $\mathbf{x}_K := \mathbf{x}(T)$. Superscript indices reference the player to which the quantity applies, e.g., $\mathbf{x}^{\scriptscriptstyle P_i}_{\scriptscriptstyle K}$ refers to the state of player $P_i$ at its final time $T$. In certain expressions we do not use the shorthand notation, e.g., $\mathbf{x}^{\scriptscriptstyle P_i}(T^{\scriptscriptstyle C})$ is the state of player $P_i$ at the final time specified for player $C$'s problem. An overbar signifies that the quantity is known and fixed in a problem formulation, and not a variable to be decided.

\section{Problem Statement}

For players $E$ and $P$ we formulate the optimal control problems: \textbf{OCP$_{\scriptscriptstyle \textbf{E}}$}, \textbf{OCP$_{\scriptscriptstyle \textbf{P}}$}, below. The goal is to determine optimal trajectories for each player that would render the opponent's problem infeasible. \\

\subsection{Optimal Control Problem for player E $(\mathbf{OCP_E})$}

Given $\{\bar{\mathbf{x}}^{\scriptscriptstyle C}, \bar{T}^{\scriptscriptstyle C}\}$, $\{\bar{\mathbf{x}}^{\scriptscriptstyle P_i}, \bar{T}^{\scriptscriptstyle P_i}\}  \  \forall \  i \in \mathcal{I} = \{1,\ldots, n\}$, and $\mathbf{x}^{\scriptscriptstyle E}(t_{\scriptscriptstyle 1})$, the following problem will produce the solution $\{\bar{\mathbf{x}}^{\scriptscriptstyle E}, \bar{\mathbf{u}}^{\scriptscriptstyle E}, \bar{T}^{\scriptscriptstyle E}\}$, if feasible.
\begin{align}
  &\minimize_{ \mathbf{x}^{\scriptscriptstyle E}, \mathbf{u}^{\scriptscriptstyle E}, T^{\scriptscriptstyle E} } \hspace{1.5cm}  \int_{t_{\scriptscriptstyle 1}}^{T^{\scriptscriptstyle E}} 1 \ dt  \label{eqn:objectiveOCPE} \\
  &\ \text{subject to} \nonumber \\
  &\phantom{\ \text{subject to}}  \ \ \ \dot{\mathbf{x}}^{\scriptscriptstyle E}(t) = \mathbf{f}^{\scriptscriptstyle E}\left( \mathbf{x}^{\scriptscriptstyle E}(t), \mathbf{u}^{\scriptscriptstyle E}(t) \right) \hspace{0.93cm} \forall \ \ t_{\scriptscriptstyle 1} \leq t \leq T^{\scriptscriptstyle E} \label{eqn:dynE} \\
  &\phantom{\ \text{subject to}} \ \ \ \mathbf{g}^{\scriptscriptstyle E}\left( \mathbf{x}^{\scriptscriptstyle E}(t), \mathbf{u}^{\scriptscriptstyle E}(t) \right) \leq \mathbf{0} \hspace{1.55cm} \forall \ \ t_{\scriptscriptstyle 1} \leq t \leq T^{\scriptscriptstyle E} \label{eqn:consE} \\
  &\textbf{(OCP$_{\scriptscriptstyle \textbf{E}}$)}  \ \ \ \ \left\lVert H\left( \mathbf{x}^{\scriptscriptstyle E}(T^{\scriptscriptstyle E}) - \bar{\mathbf{x}}^{\scriptscriptstyle C}(T^{\scriptscriptstyle E}) \right) \right\rVert_{\scriptscriptstyle 2} \leq r_c  \label{eqn:captureE} \\
  &\phantom{\ \text{subject to}}  \ \ \left\lVert H\left( \bar{\mathbf{x}}^{\scriptscriptstyle P_i}(t) - \mathbf{x}^{\scriptscriptstyle E}(t) \right) \right\rVert_{\scriptscriptstyle 2} \geq r_e \ \ \forall \ t_{\scriptscriptstyle 1} \leq t \leq \min \{T^{\scriptscriptstyle E},\bar{T}^{\scriptscriptstyle P_i}\}, \ i \in \mathcal{I}  \label{eqn:evasionE} \\
   &\phantom{\ \text{subject to}}  \ \ \ T^{\scriptscriptstyle E} \leq \bar{T}^{\scriptscriptstyle C} \label{eqn:Teconstraint}
  \end{align}
Note that the final time $T^{\scriptscriptstyle E}$ in the minimum-time objective (\ref{eqn:objectiveOCPE}) is a decision variable, implicitly defined as the capture time of Asset $C$ in constraint (\ref{eqn:captureE}), where left multiplication by matrix $H$ isolates the relative positions, and $r_c$ is the user-defined capture radius. Constraint (\ref{eqn:dynE}) represents player $E$'s dynamics while (\ref{eqn:consE}) is a general expression for any state/input constraints specific to player $E$. Constraint (\ref{eqn:evasionE}) describes the condition that $E$ must evade all pursuers with evasion radius $r_e$ until the time at which $E$ succeeds in capturing $C$ or until the pursuers no longer exist, whichever is shorter. Constraint (\ref{eqn:Teconstraint}) puts an upper bound on how much time $E$ has to capture $C$ , e.g., the opportunity to capture the Asset may exist for only a finite length of time.

\subsection{Optimal Control Problem for player P $(\mathbf{OCP_P})$} 

Given $\{\bar{\mathbf{x}}^{\scriptscriptstyle E}, \bar{T}^{\scriptscriptstyle E}\}$, and $\{\mathbf{x}^{\scriptscriptstyle P_i}(t_{\scriptscriptstyle 1})\} \  \forall \  i \in \mathcal{I} = \{1,\ldots, n\}$, the following problem will produce the solution $\{\bar{\mathbf{x}}^{\scriptscriptstyle P_i}, \bar{\mathbf{u}}^{\scriptscriptstyle P_i}, \bar{T}^{\scriptscriptstyle P_i}\} \ \forall \ i \in \mathcal{I}$, if feasible.
\begin{align}
  &\minimize_{ \{ \mathbf{x}^{\scriptscriptstyle P_i}, \mathbf{u}^{\scriptscriptstyle P_i}, T^{\scriptscriptstyle P_i} \}_{\scriptscriptstyle i=1}^{\scriptscriptstyle n} } \hspace{0.5cm}  \min_{i} \left\{ \int_{t_{\scriptscriptstyle 1}}^{T^{\scriptscriptstyle P_i}} 1 \ dt \right\}  \label{eqn:objectiveOCPP} \\
  &\ \text{subject to} \nonumber \\
  &\phantom{\ \text{subject to}}  \hspace{0.15cm} \dot{\mathbf{x}}^{\scriptscriptstyle P_i}(t) = \mathbf{f}^{\scriptscriptstyle P_i}\left( \mathbf{x}^{\scriptscriptstyle P_i}(t), \mathbf{u}^{\scriptscriptstyle P_i}(t) \right) \hspace{0.45cm} \forall \ \ t_{\scriptscriptstyle 1} \leq t \leq T^{\scriptscriptstyle P_i}, \ i=1,\ldots, n \label{eqn:dynP} \\
    &\textbf{(OCP$_{\scriptscriptstyle \textbf{P}}$)} \hspace{0.45cm} \mathbf{g}^{\scriptscriptstyle P_i}\left( \mathbf{x}^{\scriptscriptstyle P_i}(t), \mathbf{u}^{\scriptscriptstyle P_i}(t) \right) \leq \mathbf{0} \hspace{1.14cm} \forall \ \ t_{\scriptscriptstyle 1} \leq t \leq T^{\scriptscriptstyle P_i}, \ i=1, \ldots, n \label{eqn:consP} \\
  &\phantom{\ \text{subject to}} \hspace{0.2cm}  \exists \ i \in \mathcal{I}  \ni \  \left( \left\lVert H\left( \mathbf{x}^{\scriptscriptstyle P_i}(T^{\scriptscriptstyle P_i}) - \bar{\mathbf{x}}^{\scriptscriptstyle E}(T^{\scriptscriptstyle P_i}) \right) \right\rVert_{\scriptscriptstyle 2} \leq r_c \right) \wedge \left( T^{\scriptscriptstyle P_i}  < \bar{T}^{\scriptscriptstyle E} \right)  \label{eqn:captureP}
\end{align}

The expression in (\ref{eqn:objectiveOCPP}) represents an objective where the pursuer $P_i$ that can capture $E$ in the shortest amount of time decides the objective value for P as a whole. Constraints (\ref{eqn:dynP}) and (\ref{eqn:consP}) are general expressions for the pursuers' dynamics and constraints. The constraint (\ref{eqn:captureP}) describes the condition that there must exist a pursuer that can capture $E$ within a time span that is less than the time it takes $E$ to capture $C$. \\

We have captured the nature of the asset guarding game in constraints (\ref{eqn:evasionE}) and (\ref{eqn:captureP}), where the satisfaction of one constraint implies that the other cannot be satisfied. Hence, either $\mathbf{OCP_E}$  or  $\mathbf{OCP_P}$  may be feasible. Both cannot be simultaneously feasible when $\{T^{\scriptscriptstyle P_i}\}_{i=1}^{n}$ and $T^{\scriptscriptstyle E}$ take on assumed values. This fact is exploited when using the Iterative Best Response method (IBR), to be described in a later section.

\section{Trajectory Optimization using Sequential Convex Programming}

The work in \cite{Szmuk0}-\cite{Hull} describe a process of transcribing a continuous-time optimal control problem $\textbf{OCP}$ into a finite-dimensional, convex parameter optimization problem $\textbf{OPT}$, where the dynamics are first linearized about a nominal trajectory to produce a time-varying model. This continuous-time model is then discretized so that the values of state and input variables at specified temporal nodes may be solved for. Non-convex constraints on the state and input are also linearized about the trajectory to produce a convex program. A similar transcription process is completed for each of the optimal control problems stated in the \textit{Problem Statement}: $(\textbf{OCP}_{\scriptscriptstyle \textbf{E}} \rightarrow \textbf{OPT}_{\scriptscriptstyle \textbf{E}})$ and $(\textbf{OCP}_{\scriptscriptstyle \textbf{P}} \rightarrow \textbf{OPT}_{\scriptscriptstyle \textbf{P}})$. We note that $\textbf{OCP}_{\scriptscriptstyle \textbf{P}}$ may be decoupled into $n$ separate optimal control problems that are then transcribed into $n$ convex formulations, $\{\textbf{OPT}_{\scriptscriptstyle \textbf{P}_i}\}_{i=1}^n$. Each problem is solved using an SCP algorithm and we choose the best optimal value to represent $P$, i.e., $\bar{T}^P = \min \{ \bar{T}^{P_i}\}_{i=1}^n$. Since the Pursuers' problems are decoupled, we may solve them in parallel. 

When executing an SCP algorithm on a trajectory optimization problem, the approximated convex problems may become infeasible. This \textit{artificial infeasibility} \cite{Szmuk1} is frequently encountered in the early iterations of the algorithm where the dynamics are linearized about a poor initial guess. To alleviate this issue, certain implementations of SCP, such as Successive Convexification \cite{Mao1} or Penalized Trust Region \cite{Reynolds1}, add slack variables called \textit{virtual controls} to the optimization problem formulation. These virtual controls act as dynamic relaxation terms that take on nonzero values when necessary to satisfy state and input constraints. In turn, use of these slack variables is heavily penalized with a term in the objective function. As a condition for convergence in these SCP algorithms,  use of these virtual controls must become negligible at later SCP iterations so that the converged solution is dynamically feasible without the aid of these fictitious terms. Both converged (dynamically feasible) and un-converged (dynamically infeasible) SCP solutions are used in the IBR algorithm to be introduced in the following section.

\section{Iterative Best Response}

In our IBR algorithm, we use the following notation where a bold, uppercase $\mathbf{E}$ with superscript $i$ refers to player $E$'s solution at the $i^{th}$ iteration of the algorithm. Similar notation is used for player $P$'s solution at a given iteration.
\begin{equation} \label{eqn:IBRnotation}
  \mathbf{E}^i := \{ \bar{\mathbf{x}}^{\scriptscriptstyle E}, \bar{\mathbf{u}}^{\scriptscriptstyle E}, \bar{T}^{\scriptscriptstyle E} \} = 
    \begin{cases}
      \{ \bar{\mathbf{x}}_{\scriptscriptstyle 1}^{\scriptscriptstyle E}, \ldots, \bar{\mathbf{x}}_{\scriptscriptstyle k}^{\scriptscriptstyle E}, \ldots,  \bar{\mathbf{x}}_{\scriptscriptstyle K}^{\scriptscriptstyle E} \} , \\
      \{ \bar{\mathbf{u}}_{\scriptscriptstyle 1}^{\scriptscriptstyle E}, \ldots, \bar{\mathbf{u}}_{\scriptscriptstyle k}^{\scriptscriptstyle E}, \ldots,  \bar{\mathbf{u}}_{\scriptscriptstyle K}^{\scriptscriptstyle E} \} , \\
      \quad \quad \quad \bar{T}^{\scriptscriptstyle E}
    \end{cases}       
\end{equation}

The Iterative Best Response algorithm takes initial guesses for the solutions, $\mathbf{E}^0$ and $\mathbf{P}^0$, and a known trajectory $\mathbf{C}$ for the Asset. Since an asset-guarding engagement begins with the presence of the Evader $E$, we assume $E$ takes on its given initial guess trajectory and we solve for the pursuers $P$'s best response to that trajectory using the SCP subroutine on a transcribed $\mathbf{OPT_P}$. Once we obtain $P$'s best response trajectory, we then solve player $E$'s $\mathbf{OPT_E}$ to determine its best response in return. The sequential solve process for these two optimization problems using the SCP subroutine constitutes a single IBR iteration. We proceed with the iterations until a user-defined number of iterations $N_{\scriptscriptstyle IBR}$ is completed. We note that even if the SCP subroutine fails to converge for any of the players, we still use the un-converged solution and proceed with the IBR procedure. \\

\begin{algorithm}[H] \label{algo:IBR1}
    \SetKwInOut{Input}{Input}
    \SetKwInOut{Output}{Output}
    \SetKwProg{try}{try}{:}{}
    \SetKwProg{catch}{catch}{:}{end}
    \Input{$\mathbf{E}^0$, $\mathbf{P}^0$, $\mathbf{C}$}
    \Output{recorded $\mathbf{E}^i$, $\mathbf{P}^i$}
    
    \For{$i = 1 : N_{IBR}$}{

	$\mathbf{P}^i \leftarrow \mathbf{OPT_P^i} \left( \mathbf{P}^{i-1} \ ; \ \mathbf{E}^{i-1}, \ \mathbf{C} \right)$ \\
	
        \If{$ \left( \ \mathbf{P}^{i} \ \text{not converged} \ \right) \ \wedge  \ \left( \mathbf{E}^{i-1} \ \text{converged} \ \right) $}{
        record $\mathbf{E}^{i-1}$
                }
                
        $\mathbf{E}^i \leftarrow \mathbf{OPT_E^i} \left( \mathbf{E}^{i-1} \ ; \ \mathbf{P}^{i}, \ \mathbf{C} \right)$
      
              \If{$ \left( \ \mathbf{P}^{i} \ \text{converged } \right) \ \wedge  \ \left( \ \mathbf{E}^{i} \ \text{not converged } \right) $}{
        record $\mathbf{P}^{i}$
                }
    }
    \caption{Iterative Best Response for Asset Guarding Game with non-strategic Asset $C$}
\end{algorithm} 
\vspace{0.4cm}

As we proceed with the IBR iterations, we keep record of a player's converged solution if the opponent's subsequent SCP subroutine fails to converge on a solution. For instance, given that player $E$ has a converged (dynamically feasible) trajectory, if the opposing player $P$ fails to also find a converged (dynamically feasible) trajectory in response, then player $E$ has found a strategy with the potential to win the game.

\section{Numerical Examples}

For the examples in this paper, we assume point-mass-with-drag dynamics for all players. That is, the dynamics of (\ref{eqn:dynE}) and (\ref{eqn:dynP}) are replaced with (\ref{eqn:pointmassdyn}) below, where we have dropped superscripts denoting the players (i.e., $E$, $P_1, \ldots, P_n$). The state vector $\mathbf{x} \in \mathbb{R}^6$ includes a player's position $\mathbf{p} := [p_x, p_y, p_z]^\top$ and velocity $\mathbf{v} := [v_x, v_y, v_z]^\top$ in 3-dimensional space. Each player can control its motion by inducing an acceleration via its input vector $\mathbf{u} \in \mathbb{R}^3$. In addition, each body is affected by the acceleration due to atmospheric drag $\mathbf{d} \in \mathbb{R}^3$. The quantities are measured using a Cartesian coordinate system and an inertial reference frame placed at sea level with the z-axis pointing ``upwards.'' 
\begin{align} \label{eqn:pointmassdyn}
\dot{\mathbf{x}}(t) :=  \begin{bmatrix} \dot{\mathbf{p}}(t) \\ \dot{\mathbf{v}}(t) \end{bmatrix} = &\begin{bmatrix} \mathbf{v}(t) \\ \mathbf{u}(t) + \mathbf{d}\left(\mathbf{p}(t),\mathbf{v}(t)\right) \end{bmatrix}  \hspace{0.5cm} \forall \ t_{\scriptscriptstyle 1} \leq t \leq T
\end{align}

\begin{figure}[!htbp]
\centering
\includegraphics[width=1.0\textwidth]{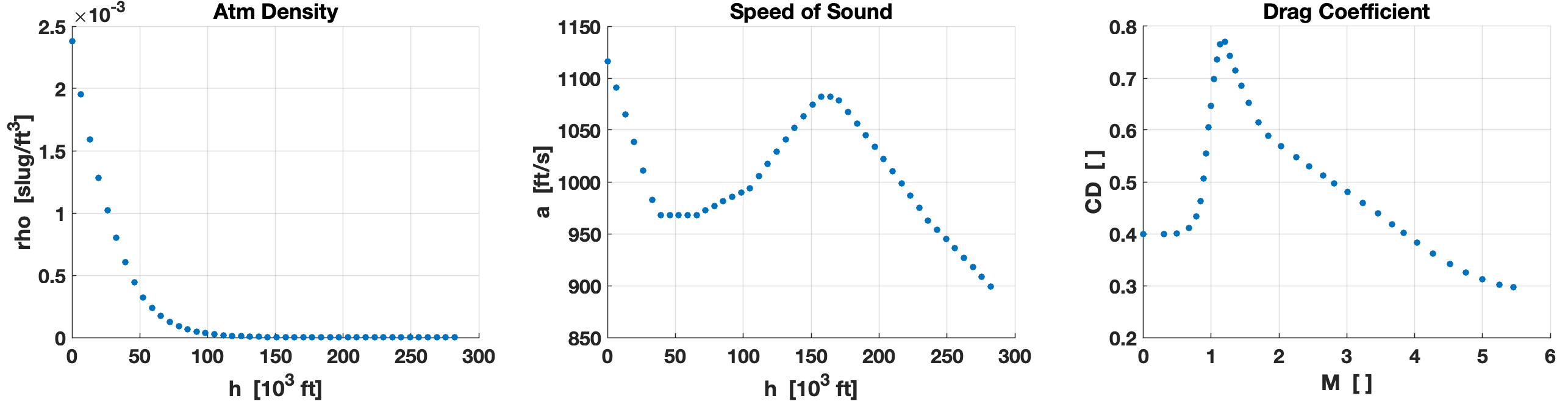}
\caption{Tabular data for atmospheric density, speed of sound, drag coefficient}
\label{fig:atmtables}
\end{figure}

\noindent The acceleration due to atmospheric drag is a function of position and velocity:
\begin{align} \label{eqn:drag}
\mathbf{d}\left( \mathbf{p}(t), \mathbf{v}(t) \right) &:= - \frac{1}{2}\frac{S}{m} \rho\left(\mathbf{p}_z(t)\right) C_D\left( M\left( \mathbf{p}(t), \mathbf{v}(t) \right) \right) \lVert \mathbf{v}(t) \rVert_{\scriptscriptstyle 2} \mathbf{v}(t)
\end{align}
where mass $m$ and reference area $S$ of the body are constant parameters. We note that atmospheric density $\rho$ is a function of altitude and that the drag coefficient $C_D$ is a function of the Mach number $M$. The Mach number depends on both the magnitude of the velocity and the speed of sound $a$, which in turn is also a function of altitude.  
\begin{align}
M\left( \mathbf{p}(t), \mathbf{v}(t) \right) &:= \frac{\lVert \mathbf{v}(t) \rVert_{\scriptscriptstyle 2}}{a\left( \mathbf{p}_z(t) \right)}
\end{align}
The atmospheric parameters ($\rho$, $a$, $C_D$) are determined using look-up tables represented in Fig. (\ref{fig:atmtables}). We use atmospheric density and speed of sound data from the 1976 U.S. Standard Atmosphere \cite{USA} and assume drag coefficient data representative of the V-2 rocket \cite{Sutton}. 

We use a piecewise cubic Hermite interpolator \cite{Fritsch} in this work as it is relatively inexpensive to evaluate but also provides continuous first derivatives. Since the dynamics and any non-convex constraints must be linearized about a nominal trajectory when using SCP, the partial derivatives of the atmospheric parameters with respect to the states must be approximated. One approach is to use the analytical partial derivative of the piecewise interpolating polynomials. Another method is to use finite differencing \cite{Nocedal} on the interpolation. In this work, we use the finite differencing approach as it provides sufficient accuracy compared to analytical partial derivatives. Furthermore, for applications where a parameter is a function of multiple states (e.g., maximum thrust as a 2D lookup table of both Mach number and altitude), finite-differencing may be more convenient than the analytical derivative approach, which requires storing and then locating polynomials on a multi-dimensional grid. 

We replace the general constraint expressions of (\ref{eqn:consE}) and (\ref{eqn:consP}) with the following box constraints on the input accelerations and a lower bound on the Mach number: 
\begin{align} 
&\lVert \mathbf{u}(t) \rVert_{\infty} \leq u_{max} \hspace{2.4cm} \forall \ t_1 \leq t \leq T \label{eqn:inputboxconstraint} \\
&M_{min} \leq M\left( \mathbf{p}(t), \mathbf{v}(t) \right) \hspace{1.4cm} \forall \ t_1 \leq t \leq T \label{eqn:minMachconstraint}
\end{align}
The lower bound on the Mach number is an important and practical constraint to include since endoatmospheric vehicles (e.g., aircraft, cruise missiles) generally have a lower bound on speed to avoid stall and maintain lift force. We note that (\ref{eqn:minMachconstraint}) is a non-convex constraint and we use the procedure described in \cite{Szmuk1} to approximate it with a convex relaxation.

In Table I, we list the engagement parameters used for all examples. We assume that a Pursuer has captured the Evader (or the Evader has captured the Asset) if their relative distance falls under 1 [ft]. We enforce the Evader to maintain a comfortable 500 [ft] distance from the Pursuers to avoid capture. All players have a mass of 1,000 [slugs] and a cross-sectional diameter of 5 [ft], representative of the V-2 rocket. Each player's Mach number is constrained to stay above 0.5, or about 500 [ft/s].
\begin{table}[!htbp]
\centering
\begin{tabular}{@{}lllll@{}} \toprule
 Parameter & Value & Units & Description & First Mention \\ \midrule
\ \ \ $r_c$          & 1 & [ft] & Capture Radius & Eqns  (\ref{eqn:captureE}), (\ref{eqn:captureP})  \\
\ \ \ $r_e$          & 500 & [ft] & Evasion Radius & Eqn  (\ref{eqn:evasionE})  \\
\ \ \ $m$            & 1,000 & [slugs] & Mass & Eqn  (\ref{eqn:drag})    \\
\ \ \ $S$             & $\frac{\pi}{4}(5)^2$ & [ft$^2$] & Reference Area & Eqn  (\ref{eqn:drag})    \\ 
\ \ \ $M_{min}$  & 0.5  & [ \ ] & Minimum Mach Number & Eqn  (\ref{eqn:minMachconstraint})    \\ 
\ \ \ $G$      & 32.174 & [ft/s$^2$] & Standard Gravity \\ \bottomrule
\end{tabular}
\caption{Engagement Parameters}
\end{table}

\begin{figure}[!htbp]
\centering
\includegraphics[width=1.0\textwidth]{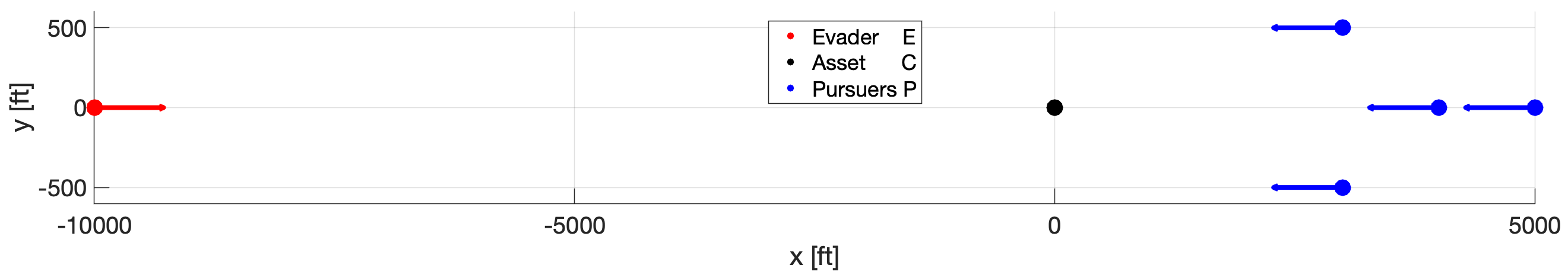}
\caption{ Initial positions and velocity vectors }
\label{fig:ICs}
\end{figure}

We study four example engagements, each subsequent example including an additional Pursuer, i.e., Example 1 includes $E$ versus $P_1$, Example 2 includes $E$ versus \{$P_1$, $P_2$\}, Example 3 includes $E$ versus \{$P_1$, $P_2$, $P_3$\} and Example 4 includes $E$ versus \{$P_1$, $P_2$, $P_3$, $P_4$\}.
The initial positions and velocity vectors are depicted with a birds-eye view in Fig. \ref{fig:ICs} where the Evader starts 1,000 [ft] above the Asset and Pursuers. The direction of the velocity vectors imply head-on engagements (as opposed to a tail-chase). We verify by simulation that these initial conditions describe a well-posed engagement where each player can nominally intercept its target using the proportional navigation (PN) guidance law with navigation ratio $N'=3$ \cite{Zarchan}.
As described in Table II, the Asset is assumed to remain stationary while the Evader and Pursuers begin with the same speed. We restrict the Evader's input acceleration to be 1[G] less in magnitude compared to that of the Pursuers.

\begin{table}[!htbp]
\centering
\begin{tabular}{@{}llllllll@{}} \toprule
 Player & $p_x$[ft]   & $p_y$[ft] & $p_z$[ft]  & $v_x$[ft/s]  & $v_y$[ft/s] & $v_z$[ft/s] & $u_{max}$[G] \\ \midrule
\ \ \ $E$    & -10,000 & \phantom{-50}0    & 31,000 & \phantom{0-}3,000  & \phantom{000}0    & \phantom{000}0 &  \phantom{00}7.0   \\
\ \ \ $C$    & \phantom{-10,00}0      & \phantom{-50}0    & 30,000 & \phantom{-1,000}0     & \phantom{000}0    & \phantom{000}0 &  \phantom{000}- \\
\ \ \ $P_1$ & \phantom{-1}4,000   & \phantom{-50}0    & 30,000 & \phantom{0}-3,000 & \phantom{000}0    & \phantom{000}0 &  \phantom{00}8.0   \\
\ \ \ $P_2$ & \phantom{-1}5,000   & \phantom{-50}0    & 30,000 & \phantom{0}-3,000 & \phantom{000}0    & \phantom{000}0 &  \phantom{00}8.0   \\
\ \ \ $P_3$ & \phantom{-1}3,000   & -500 & 30,000 & \phantom{0}-3,000 & \phantom{000}0    & \phantom{000}0 &  \phantom{00}8.0   \\
\ \ \ $P_4$ & \phantom{-1}3,000   & \phantom{-}500  & 30,000 & \phantom{0}-3,000 & \phantom{000}0    & \phantom{000}0 &  \phantom{00}8.0   \\   \bottomrule
\end{tabular}
\caption {Initial conditions and acceleration bounds, i.e.,  $\lVert \mathbf{u} \rVert_{\scriptscriptstyle \infty} \leq u_{\scriptscriptstyle max}$}
\end{table}

The IBR-SCP method requires an initial solution guess that consists of the state and input trajectories of all players. We may use the PN-guided trajectories described above, however, we have observed that crude straight-line state trajectories and zero input trajectories converge to trajectories similar to those warm-started by PN-based initial guesses within one IBR iteration.

After running the IBR-SCP solution method for each of the examples, we verify the solutions in simulation against opponents using conventional guidance laws, i.e., PN and Augmented PN laws with navigation ratios $N'=3,4,5$. For example, if the algorithm's result implies that the Evader successfully evades its Pursuers and reaches the Asset, then we pit the Evader's open-loop strategy against six instances of each Pursuer in simulation (two guidance laws, three navigation ratio values). This verification step checks if the solution provided by the IBR-SCP method is indeed feasible for the optimal control problem formulations listed in the \textit{Problem Statement}.

\section{Results}

\begin{figure}[!htbp]
\centering
\includegraphics[width=1.0\textwidth]{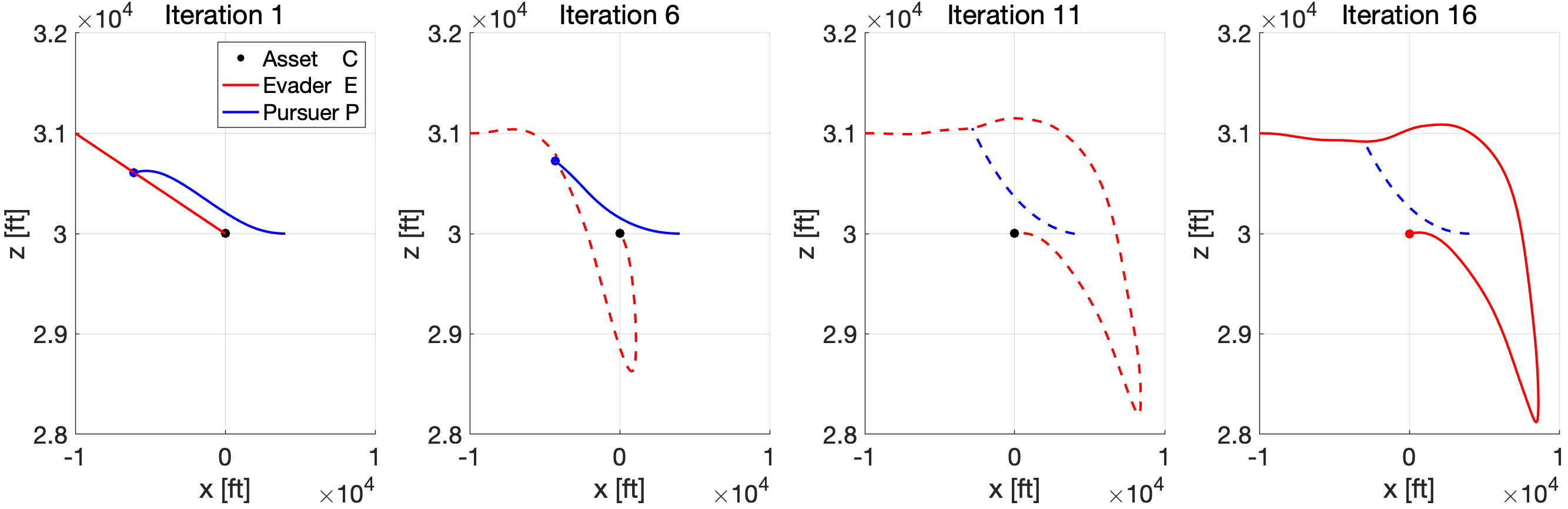}
\caption{ Trajectories for $E$ and $P_1$ evolve after each IBR iteration }
\label{fig:iterations}
\end{figure}

\begin{figure}[!htbp]
\centering
\includegraphics[width=1.0\textwidth]{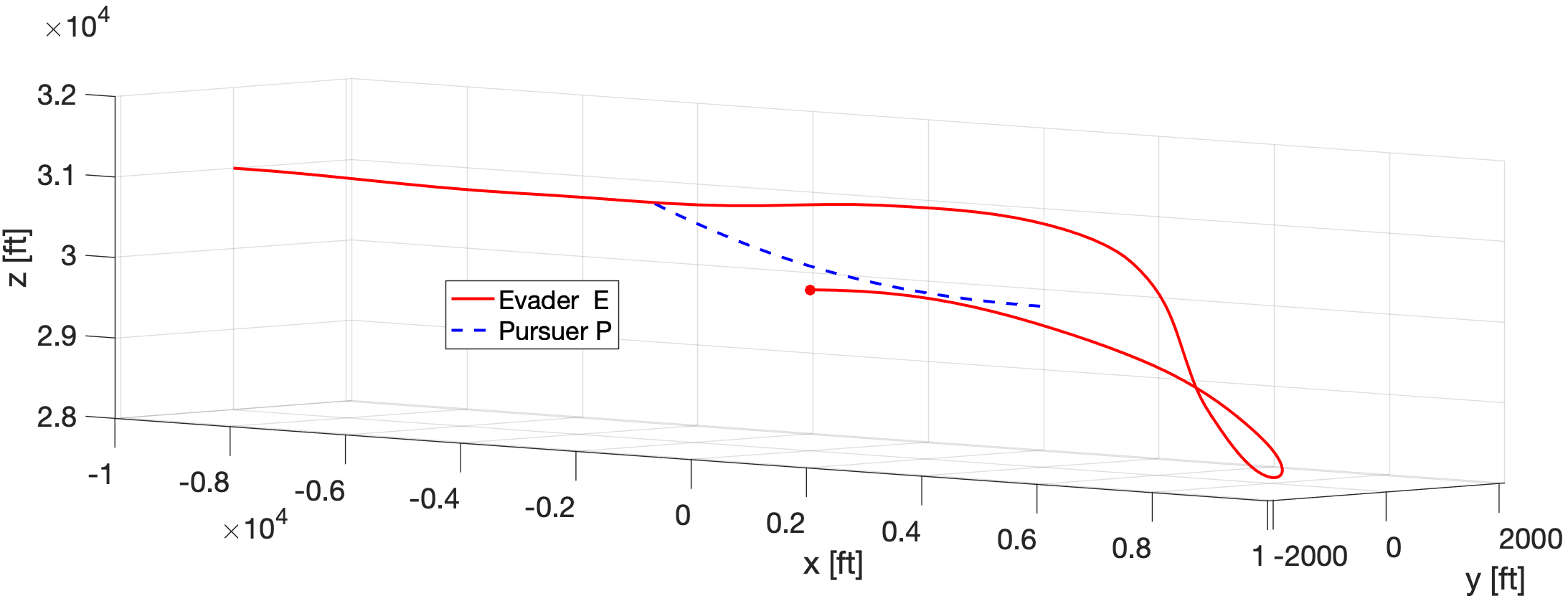}
\caption{3-D trajectories for $E$ and $P_1$ after final IBR iteration}
\label{fig:solution}
\end{figure}

\begin{figure}[!htbp]
\centering
\includegraphics[width=1.0\textwidth]{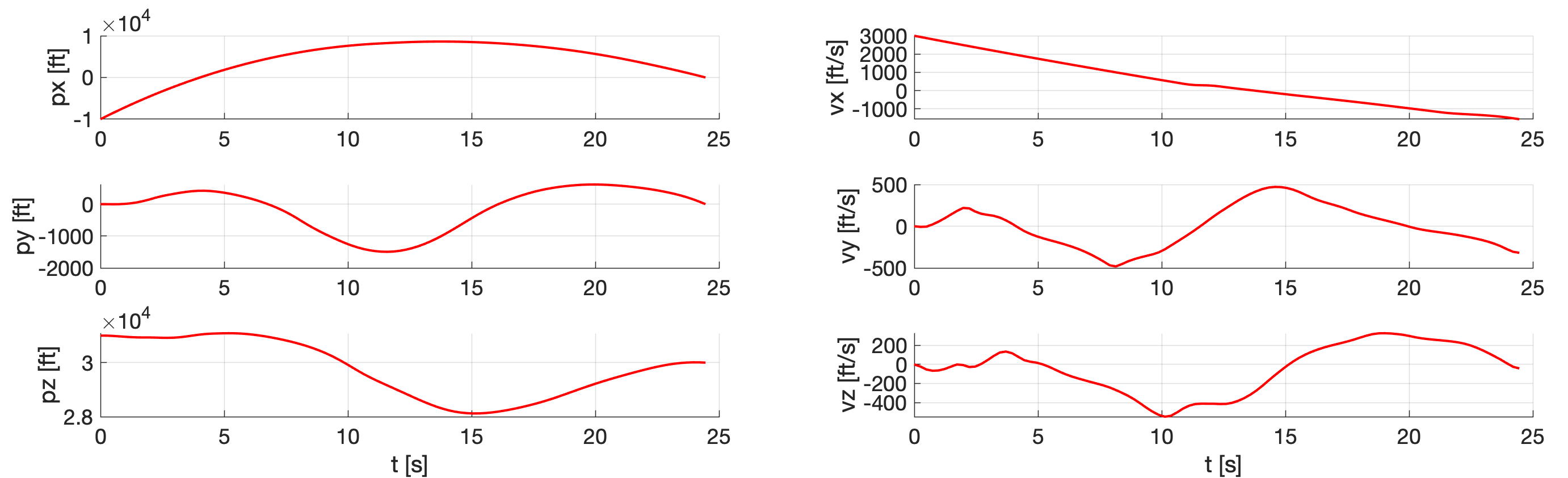}
\caption{Evader's position and velocity trajectories}
\label{fig:posvelplots}
\end{figure}

\begin{figure}[!htbp]
\centering
\includegraphics[width=1.0\textwidth]{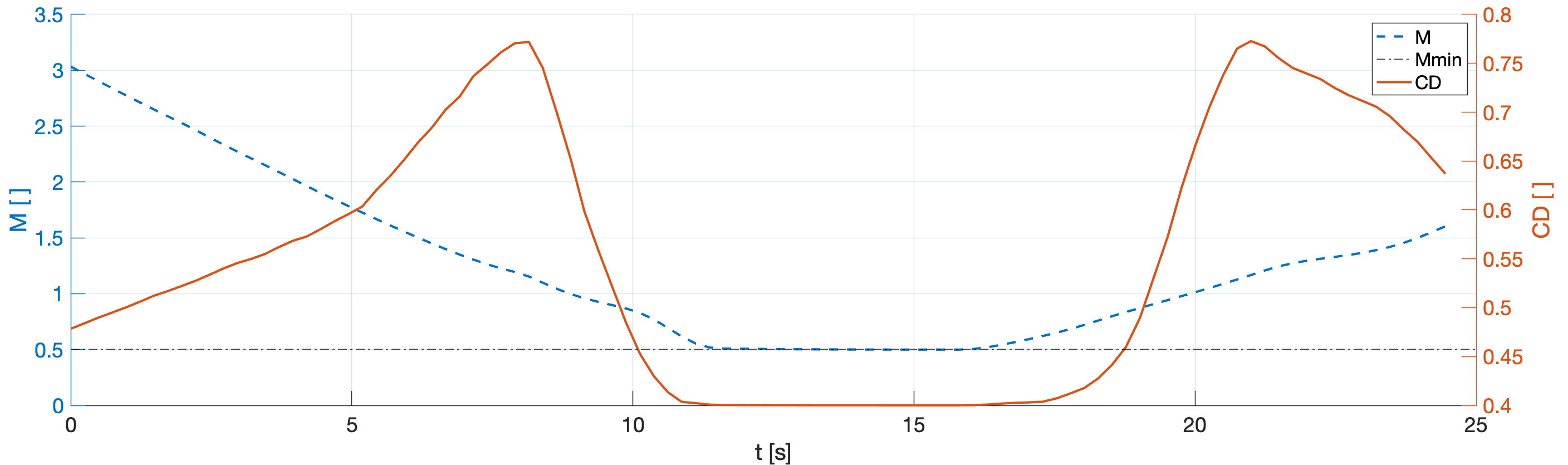}
\caption{Evader's Mach Number and Drag Coefficient}
\label{fig:MachCDplots}
\end{figure}

\begin{figure}[!htbp]
\centering
\includegraphics[width=1.0\textwidth]{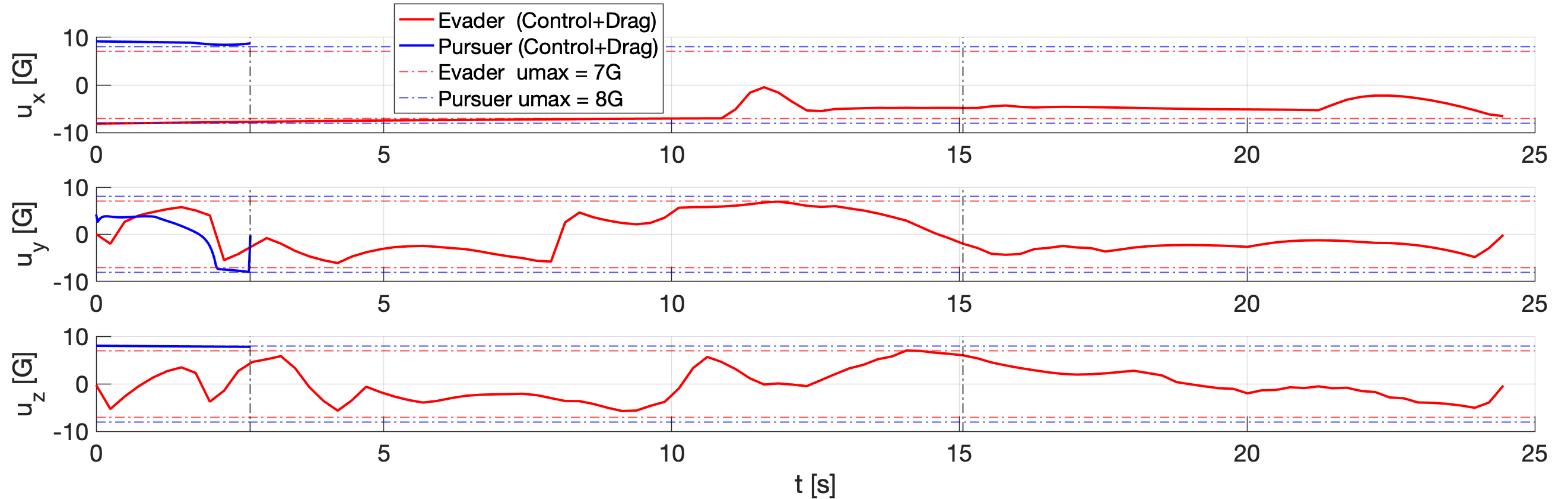}
\caption{Effective acceleration of Evader $E$ and Pursuer $P_1$}
\label{fig:accelplots}
\end{figure}

\begin{figure}[!htbp]
\centering
\includegraphics[width=0.8\textwidth]{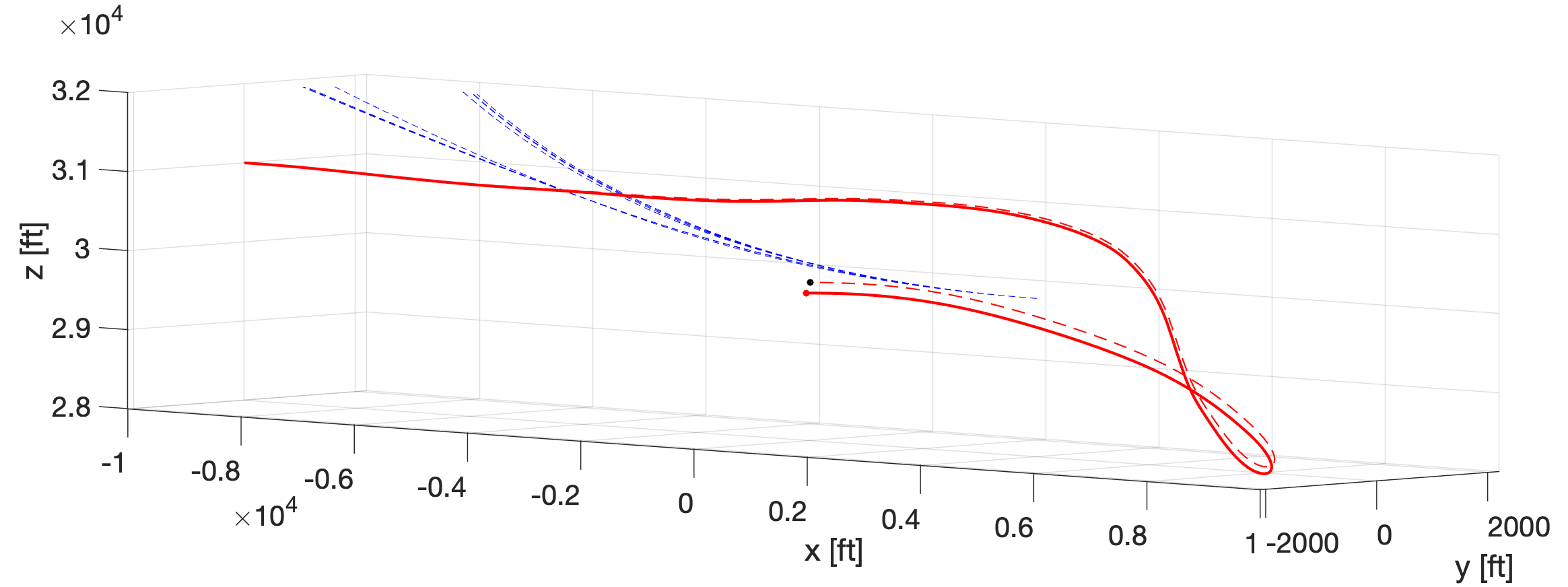}
\includegraphics[width=0.8\textwidth]{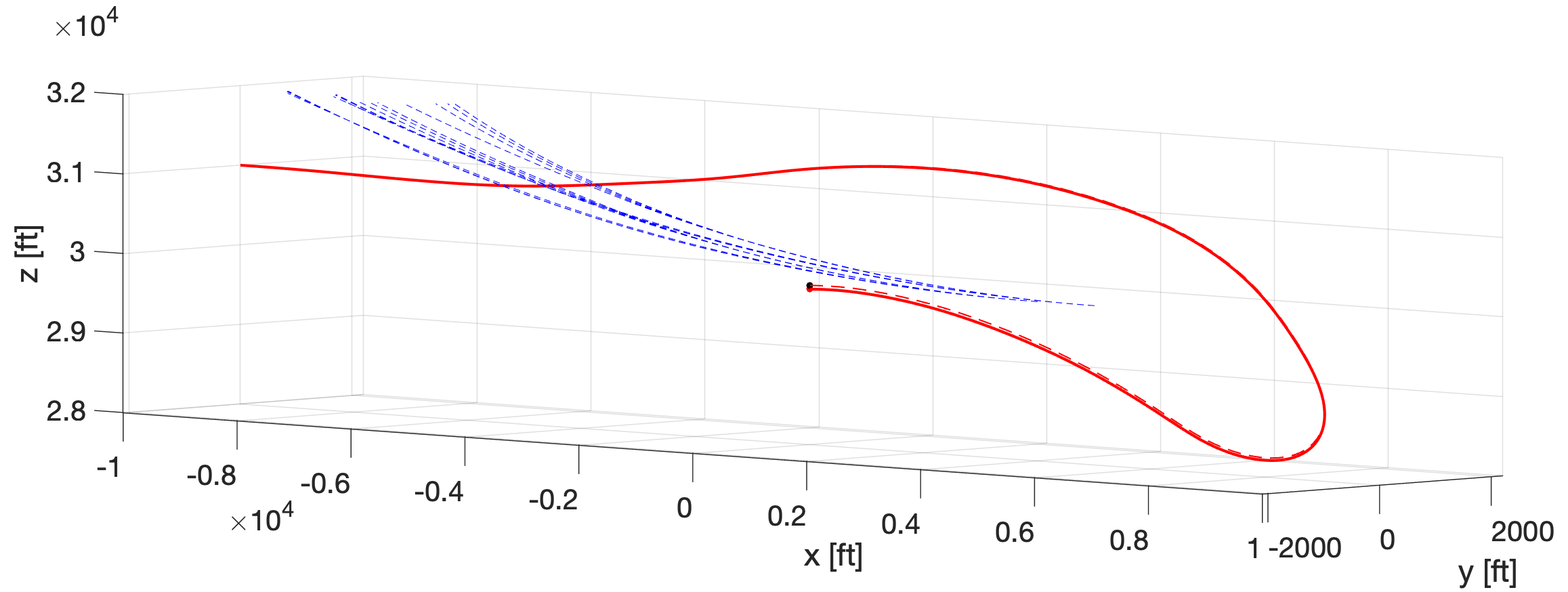}
\includegraphics[width=0.8\textwidth]{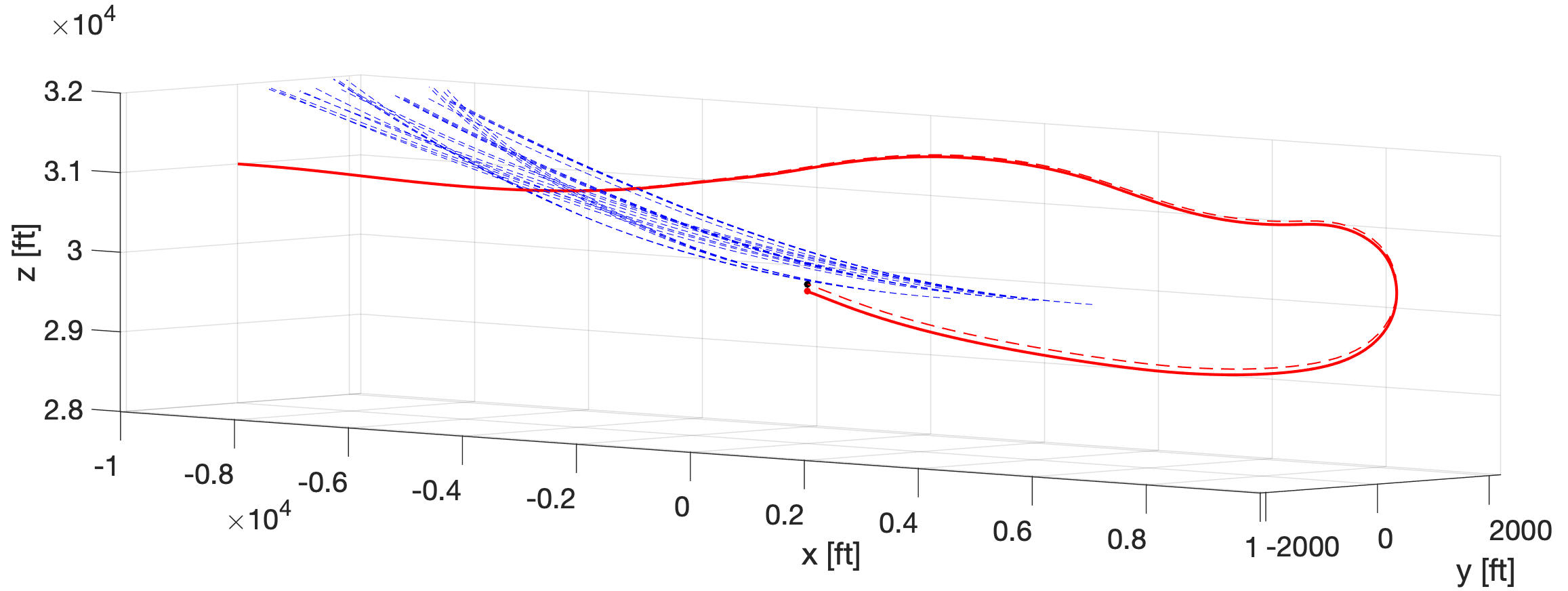}
\includegraphics[width=0.8\textwidth]{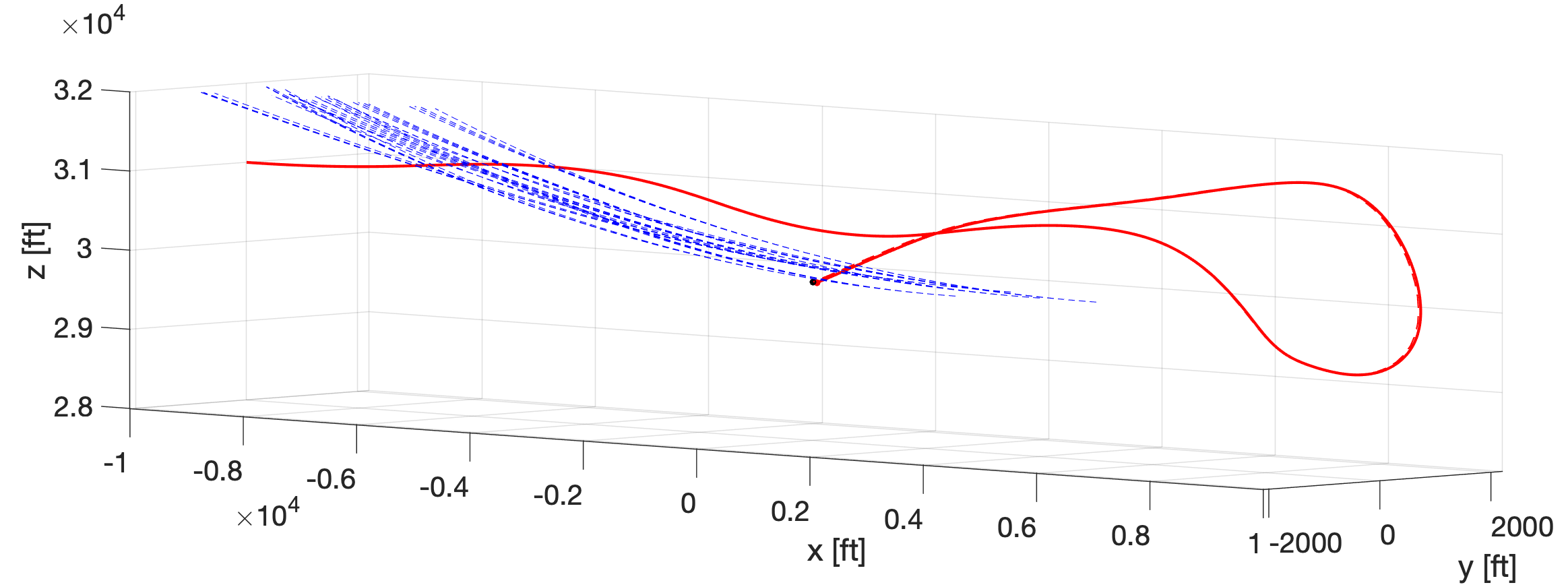}
\caption{Top to bottom: Evader (red, solid open-loop trajectory, dashed predicted trajectory) successfully evades against 1-, 2-, 3-, 4-Pursuers (blue, dashed PN law guided trajectories)}
\label{fig:verification}
\end{figure}

We apply the IBR-SCP solution method to the four asset-guarding game examples described in the prior section. Figures \ref{fig:iterations}-\ref{fig:accelplots} pertain to Example 1 ($E$ versus $P_1$) where we discuss the solution in depth. The final figure \ref{fig:verification} shows simulation results for all four examples where we verify the open-loop Evader trajectories against PN-guided Pursuer trajectories.

In Fig. \ref{fig:iterations}, we show a subset of twenty IBR iterations we run for Example 1. In the first IBR iteration we observe that the SCP subroutine has converged on a blue Pursuer trajectory that intercepts the initial straight-line guess provided for the red Evader. After five IBR iterations, we observe that the Evader's SCP subroutine has failed to converge on a dynamically feasible solution, signaled by the dashed line. Nevertheless, we solve for the Pursuer's response to that trajectory and find that it can still intercept the Evader's artificial trajectory. After five more IBR iterations we observe that neither the Evader's nor the Pursuer's SCP subroutines have converged on dynamically feasible solutions. The trajectories still intercept but are only feasible given the virtual controls that the SCP subroutine employs. After a few more IBR iterations we find a solution where the Evader is dynamically feasible but the Pursuer is not. We continue until the twentieth IBR iteration where we observe that the Evader's state trajectory has not changed significantly from that shown in IBR Iteration 16.

Fig. \ref{fig:solution} shows a scaled plot of the 3-D position trajectories found at the last IBR iteration, where it is appears that the Evader is intercepted mid-course by the Pursuer. However we recall that the blue dashed line implies that this Pursuer trajectory is not dynamically feasible. The Evader then dives in altitude and performs a ``turn-around'' maneuver to reach the Asset at the red point. We expect that if the Evader commits to the open-loop strategy described by this solution, then the Pursuer will be unable to intercept the Evader and defend the Asset. The Evader's state trajectories are shown in Fig. \ref{fig:posvelplots} where we observe that the Evader reaches the Asset's location in 24.45 seconds with a terminal speed of 1,593.1 [ft/s]. 

Fig. \ref{fig:MachCDplots} plots the Evader's Mach number on the left-hand axis and the drag coefficient on the right-hand side axis as functions of time. Considering the Evader's minimum-time objective we would expect it to maintain a high Mach number (or speed) throughout the engagement. Counterintuitive to our expectation, the Evader maintains the lowest possible Mach number for a significant duration of the engagement. This behavior is attributable to the turnaround maneuver that the Evader conducts after evading the Pursuer. Rather than taking a large-radius turn back to the Asset, the minimum-time strategy is to reduce speed drastically so that it may turn more sharply into the direction of its target. This observation is in line with minimum-time turn maneuvers by supersonic aircraft \cite{Hedrick}. In fact, when we remove the lower bound constraint on the Mach number and run the IBR algorithm, we find that the Evader reduces its speed significantly from an initial value of 3,000 [ft/s] down to approximately 60 [ft/s] to conduct a high-agility turn maneuver. Although interesting to observe in simulation, such a maneuver may result in loss of lift for conventional aircrafts. In the same plot we note that the Evader endures the penalty of a high drag coefficient as it reduces its Mach number below the drag-divergence value of 1.2. After completing the turnaround maneuver, the Evader speeds up and once again passes through the high drag coefficient regime to reach the Asset in minimum-time.

In Fig. \ref{fig:accelplots} we show the effective (control and drag) accelerations profiles for both the Evader and the Pursuer. The Pursuer's predicted accelerations stop at 2.67 seconds, when we infer that the Evader has successfully evaded the Pursuer. Considering the initial high speeds of the players in this head-on engagement, the players' strategies are to apply maximum deceleration in their respective attempts to evade and capture. The ``hard-braking'' by the Evader in the x-axis direction forces the Pursuer to follow suit in its direction of motion. We note that both players actually surpass their acceleration bounds by optimally taking advantage of atmospheric drag. In the y-axis and z-axis directions we observe that the Evader's strategy causes the Pursuer to hit and ride its acceleration bounds. The Evader has found a strategy that forces the Pursuer to apply maximum effort with all of its ability, and yet we know that the Pursuer's trajectory is artificially supported (since the SCP subroutine failed to converge on a solution that has weaned itself off of virtual controls).

In all remaining Examples 2-4, where  we add pursuers to the engagement, the IBR algorithm finds successful Evader trajectories by the $17^{\text{th}}$ iteration. Using the notation introduced in (\ref{eqn:IBRnotation}), we may measure the change in the Evader's solution after each iteration by evaluating the Frobenius norm of the relative state trajectories between iterations: $\lVert \{\bar{\mathbf{x}}^{\scriptscriptstyle E} \}^{i} - \{\bar{\mathbf{x}}^{\scriptscriptstyle E} \}^{i-1} \rVert_F$. If this quantity falls under some tolerance value $\epsilon_{\scriptscriptstyle IBR}$ and remains under for subsequent iterations, we may consider this as a condition for IBR algorithm convergence. In each of the four examples we find that this quantity takes on large values in early iterations but drops significantly to stay under $\epsilon_{\scriptscriptstyle IBR}=$1e-2 by the IBR iteration numbers listed in Table III.  With this definition of convergence and tolerance value, we may say that the IBR algorithm converges on Evader solutions for all four examples by the $19^{\text{th}}$ iteration.
\begin{table}[!htbp]
\centering
\begin{tabular}{@{}lllll@{}} \toprule
 Example & \hspace{0.6cm}Players & IBR Iteration $i$ & \quad $\lVert \{\bar{\mathbf{x}}^{\scriptscriptstyle E} \}^{i} - \{\bar{\mathbf{x}}^{\scriptscriptstyle E} \}^{i-1} \rVert_F$ \\ \midrule
\ \ \ \ 1 & \ \ \ \ \ $E$ vs. $P_1$     & \quad \quad \quad 16 & \quad \quad \quad \quad 0.06e-2  \\
\ \ \ \ 2 & \ \ \ \ \ $E$ vs. \{$P_1$,$P_2$\}     & \quad \quad \quad 12 & \quad \quad \quad \quad 0.18e-2   \\
\ \ \ \ 3 & \ \ \ \ \ $E$ vs.  \{$P_1$,$P_2$,$P_3$\}    & \quad \quad \quad 19 & \quad \quad \quad \quad 0.05e-2    \\
\ \ \ \ 4 & \ \ \ \ \ $E$ vs. \{$P_1$,$P_2$,$P_3$,$P_4$\}   & \quad \quad \quad 13 & \quad \quad \quad \quad 0.35e-2    \\ 
\end{tabular}
\caption{In each example, the change in Evader's state trajectory is minimal in later IBR iterations, implying IBR algorithm convergence}
\end{table}
\pagebreak

These four examples highlight the ability to find successful Evader strategies. For different initial conditions we do find situations where the IBR algorithm converges on solutions where the Pursuers are successful. Furthermore, there are initial conditions where the algorithm does not converge on a solution within the user-defined number of IBR iterations $N_{\scriptscriptstyle IBR}$.

Fig. \ref{fig:verification} illustrates our trajectory verification step (described in the prior \textit{Numerical Examples} section) where the Evader's input trajectories found via the IBR-SCP method are applied as open-loop strategies in simulation against Pursuers guided by conventional guidance laws. In all four examples, the (solid, red) simulated Evader trajectories are successful in evading the (dashed, blue) Pursuers guided by either PN or APN laws. To illustrate, we consider the fourth plot in Fig. \ref{fig:verification} and observe that the Evader trajectory avoids all twenty-four instances of the Pursuers (four pursuers, two guidance laws, three navigation ratios).

We note that the (solid, red) open-loop, simulated Evader trajectories do not exactly match the (dashed, red) predicted trajectories found by the IBR-SCP method (see first plot of Fig. \ref{fig:verification}). We attribute this error to the various approximations made by the SCP subroutine. Using the linear, time-varying model corresponding to the trajectory found at the last IBR iteration (produced as a byproduct of the SCP method), we design a finite-horizon, discrete-time LQR tracking control law \cite{Anderson} to be used online. By choosing the terminal state weights to be relatively larger than the running state and input weights, we produce closed-loop, simulated Evader trajectories that not only evade Pursuers but also reach the Asset. Recall that we had buffered the Evader's input acceleration to be 1[G] less in magnitude compared to that of the Pursuers in the IBR-SCP method (see Table II). The acceleration buffer may be used to realize corrective control actions produced by the feedback law.

\section{Conclusion}
 
We have implemented an Iterative Best Response (IBR) algorithm to find solutions in a multi-body asset guarding game. In particular, we have found open-loop trajectories for an Evader to evade multiple, more-maneuverable Pursuers and reach its target in minimum-time given state and input constraints. We capture the nature of the game by formulating a set of constrained optimal control problems that are coupled based on the roles taken by players in the game. The generalized modeling framework allows us to consider a wide variety of applications that can be modeled as asset-guarding games, with an arbitrary number of Pursuers, different dynamical models, objectives and constraints. Leveraging recent advances in Sequential Convex Programming (SCP), we efficiently perform constrained trajectory optimization for the players as a subroutine in the IBR solution method. The players may be modeled as nonlinear dynamical systems with data in tabular form, opening up the potential to apply the IBR-SCP solution method to a wide class of practical applications. By implementing a tracking feedback control law about the solution trajectories that we find, we may mitigate the effects of model mismatch caused by the approximations used in the approach. Future work will apply the IBR-SCP solution method to other, more general differential games. In doing so, we will also attempt to find and characterize different solutions, including Nash equilibria. Furthermore, efforts should be made to understand the convergence properties of the algorithm and determine if any guarantees can be made. 

\section{acknowledgements}
The authors gratefully acknowledge support from the Office of Naval Research under grant N00014-18-1-2209. We thank our collaborators Jyot Buch from the University of Michigan, and Kate Schweidel, Alex Devonport, He Yin at the University of California, Berkeley.


\end{document}